# Existence domains of arbitrary amplitude nonlinear structures in two-electron temperature space plasmas. II. High-frequency electron-acoustic solitons

S. K. Maharaj,[1] R. Bharuthram,[2] S. V. Singh,[3,4] and G. S. Lakhina[3]
[1]*South African National Space Agency (SANSA) Space Science, P.O. Box 32, Hermanus 7200, South Africa*
[2]*University of the Western Cape, Modderdam Road, Bellville 7530, South Africa*
[3]*Indian Institute of Geomagnetism, New Panvel (West), Navi Mumbai 410218, India*
[4]*School of Chemistry and Physics, University of KwaZulu-Natal, Private Bag X54001, Durban 4000, South Africa*

A three-component plasma model composed of ions, cool electrons, and hot electrons is adopted to investigate the existence of large amplitude electron-acoustic solitons not only for the model for which inertia and pressure are retained for all plasma species which are assumed to be adiabatic but also neglecting inertial effects of the hot electrons. Using the Sagdeev potential formalism, the Mach number ranges supporting the existence of large amplitude electron-acoustic solitons are presented. The limitations on the attainable amplitudes of electron-acoustic solitons having negative potentials are attributed to a number of different physical reasons, such as the number density of either the cool electrons or hot electrons ceases to be real valued beyond the upper Mach number limit, or, alternatively, a negative potential double layer occurs. Electron-acoustic solitons having positive potentials are found to be supported only if inertial effects of the hot electrons are retained and these are found to be limited only by positive potential double layers. © 2012 American Institute of Physics. [http://dx.doi.org/10.1063/1.4769174]

## I. INTRODUCTION

It is well known that two electron populations having markedly different temperatures are quite common in space plasmas. Such plasmas can support the propagation of the linear electron-acoustic wave where the restoring force is provided by the hot electrons but the inertial effects required to sustain the oscillations are provided by the cooler of the two electron species.

The importance of beam-driven linear electron-acoustic wave instabilities in two-electron temperature plasmas in explaining high-frequency broadband electrostatic noise (BEN) below the local electron plasma frequency has been well established.[1–3] The higher frequency part of the dayside auroral region BEN having frequencies exceeding the local electron plasma frequency has been attributed to electron-acoustic solitons which evolve from the nonlinear stage of the electron-acoustic instability.[4,5] The existence of both small and large amplitude electron-acoustic solitons has been theoretically investigated by Mace *et al.*[6] for a plasma composed of Boltzmann hot electrons, cool ions, and cool electrons, where both cool species are assumed to be adiabatic fluids. They found only negative potential electron-acoustic solitons.

A number of observations by satellites such as measurements by FAST in the downward current region of the mid-altitude auroral region,[7] by POLAR in the high altitude polar magnetosphere,[8] by GEOTAIL in the magnetotail region,[9] by Wind in the terrestrial bow shock,[10] and, more recently, by CLUSTER in the dayside magnetosheath,[11] reveal the presence of travelling bipolar electric-field structures in the component parallel to the magnetic field. The observed electrostatic solitary waves (ESWs) were found to have positive potentials and seem to be associated with high frequency electron dynamics. This apparently ruled out the possibility of interpreting the ESWs in terms of electron-acoustic solitons or double layers. A satisfactory theoretical explanation on the basis of Bernstein-Greene-Kruskal theory was that the observed ESWs could be interpreted as localized electrostatic potential perturbations associated with a moving bunch of trapped electrons.[12]

An important breakthrough in the theory of nonlinear electron-acoustic waves was that large amplitude electron-acoustic solitons having positive polarity are possible when the two-electron component model composed of hot electrons which are inertialess and inertial cool electrons[4–6] was extended by Berthomier *et al.*[13] to three-electron components by including an additional component of inertial beam electrons. Later on, theoretical investigations of Cattaert *et al.*,[14] based on the fluid-dynamic paradigm approach (Refs. [15–17]), revealed that positive potential electron-acoustic solitons are possible even for a two-electron (cold and hot) model provided that the inertia of the hot electrons is also included in the analysis. The existence domains for electron-acoustic solitons corresponding to the values $\gamma = 1$ (Boltzmann case) and $\gamma = 3$ (adiabatic case) for the polytropic index of the hot electrons were found to be qualitatively quite similar to the $\gamma = 2$ case.

Reverting to nonlinear studies which use the traditional Sagdeev approach, the existence of arbitrary amplitude ion-acoustic and electron-acoustic solitons was investigated by Lakhina et al.[18] for a three-component plasma model composed of ions, and cool and hot electrons, where all species were treated as adiabatic fluids. The parameter regions which support the occurrence of large amplitude ion-acoustic and electron-acoustic solitons have been identified; however, the key finding was that inclusion of inertial effects of the hot electrons in the model allowed for positive potential electron-acoustic solitons to be supported which is consistent with the findings in Ref. 14. Only the lower limits of the Mach number ranges which support ion-acoustic and electron-acoustic solitons were calculated in Ref. 18. Although it was only briefly pointed out in Ref. 18 that upper Mach number limits exist for ion-acoustic and electron-acoustic solitons, reasons were not provided as to why these upper Mach number limits for solitons should occur.

The existence of large amplitude electron-acoustic solitary waves for a model composed of inertial cool electrons, inertialess hot electrons which have a kappa velocity distribution and ions was investigated by Danehkar et al.[19] and Devanandhan et al.[20] Only negative potential structures were found for which the upper Mach number limitations as discussed in Ref. 19 were found to be imposed by the number density of the cool electrons becoming complex valued. Recently, Singh et al.[21] have discussed electron-acoustic solitons in a four-component model composed of ions, cool electrons, beam electrons (all three species are inertial), and non-thermal[22] hot electrons which are inertialess. Non-thermal effects of the hot electrons were found to reduce the amplitude of electron-acoustic solitons having negative potentials. For hot electrons which are Boltzmann distributed, the existence domains for negative as well as positive potential electrostatic solitons and double layers were obtained. The pattern of the existence domains of large amplitude electron-acoustic solitary waves was found to be very similar not only to that found by Cattaert et al.[14] but also to the existence pattern of dust-acoustic solitons found by Verheest et al.[23] for a dusty plasma model composed of cold and adiabatic dust (both negatively charged) and Boltzmann ions and Boltzmann electrons which are both very hot.

This paper is a continuation of our work reported earlier by Maharaj et al.,[24] henceforth referred to as I, where the existence of large amplitude ion-acoustic solitons was discussed for the three-component plasma model comprised ions, cool electrons, and hot electrons, both for the model of Lakhina et al.[18] which assumed that all species are inertial and adiabatic and the model of Mace et al.[6] which did not take into consideration inertial effects of the hot electrons, whilst still retaining inertia and pressure for the ions and cool electrons. Here, the focus is on the high-frequency regime and we discuss the existence of large amplitude electron-acoustic solitons. We are interested primarily in why upper Mach number limits occur for electron-acoustic solitons and we explicitly calculate these upper limiting values of the Mach number. We consider much broader regions in parameter space compared to those discussed by Lakhina et al.,[18] and present here the permitted Mach number ranges supporting the existence of electron-acoustic solitons having taken into consideration both the lower and upper Mach number limits.

This paper is organised as follows. In Sec. II, we present the model of Lakhina et al.[18] composed of ions, cool and hot electrons for which inertia and pressure have been retained for all species. The model of Mace et al.[6] which does not take into consideration inertial effects of the hot electrons but includes the inertia and pressure of the ions and cool electrons is presented in Sec. III. Details of the theory are omitted in Secs. II and III since these were included in the relevant sections in I (Ref. 24). For completeness, we include in each of Secs. II and III the final expression for the Sagdeev potential and the relevant expressions for the number densities of the different species which were used to obtain the final expression. In Sec. IV, we present numerical results and discussion. Our findings are summarized in Sec. V.

## II. MODEL WHICH INCLUDES HOT ELECTRON INERTIA

The model comprised of ions, and cool and hot electron components where all species are treated as adiabatic fluids[18] is discussed in Sec. II in I. The expressions for the number density of the ions, cool electrons, and hot electrons as given in I, respectively, are given by

$$n_i = \frac{1}{2\sqrt{3}}\{[(M+\sqrt{3})^2 - 2\Phi]^{1/2} - [(M-\sqrt{3})^2 - 2\Phi]^{1/2}\}, \quad (1)$$

$$n_{ce} = \frac{n_{ce}^0}{2\sqrt{3T_{ce}/\mu_e}}\{[(M+\sqrt{3T_{ce}/\mu_e})^2 + (2\Phi/\mu_e)]^{1/2} - [(M-\sqrt{3T_{ce}/\mu_e})^2 + (2\Phi/\mu_e)]^{1/2}\}, \quad (2)$$

and

$$n_{he} = \frac{n_{he}^0}{2\sqrt{3T_{he}/\mu_e}}\{[(M+\sqrt{3T_{he}/\mu_e})^2 + (2\Phi/\mu_e)]^{1/2} - [(M-\sqrt{3T_{he}/\mu_e})^2 + (2\Phi/\mu_e)]^{1/2}\}, \quad (3)$$

which, upon substitution in Poisson's equation, yields the expression for the Sagdeev potential given by

$$V(\Phi, M) = \frac{1}{6\sqrt{3}}\left\{(M+\sqrt{3})^3 - \left(\sqrt{(M+\sqrt{3})^2 - 2\Phi}\right)^3\right\} - \frac{1}{6\sqrt{3}}\left\{(M-\sqrt{3})^3 - \left(\sqrt{(M-\sqrt{3})^2 - 2\Phi}\right)^3\right\}$$

$$+ \frac{n_{ce}^0}{6\sqrt{3T_{ce}/\mu_e}}\mu_e\left\{(M+\sqrt{3T_{ce}/\mu_e})^3 - \left(\sqrt{(M+\sqrt{3T_{ce}/\mu_e})^2 + \frac{2\Phi}{\mu_e}}\right)^3\right\}$$

$$- \frac{n_{ce}^0}{6\sqrt{3T_{ce}/\mu_e}}\mu_e\left\{(M-\sqrt{3T_{ce}/\mu_e})^3 - \left(\sqrt{(M-\sqrt{3T_{ce}/\mu_e})^2 + \frac{2\Phi}{\mu_e}}\right)^3\right\}$$

$$+ \frac{n_{he}^0}{6\sqrt{3T_{he}/\mu_e}}\mu_e\left\{(M+\sqrt{3T_{he}/\mu_e})^3 - \left(\sqrt{(M+\sqrt{3T_{he}/\mu_e})^2 + \frac{2\Phi}{\mu_e}}\right)^3\right\}$$

$$- \frac{n_{he}^0}{6\sqrt{3T_{he}/\mu_e}}\mu_e\left\{(M-\sqrt{3T_{he}/\mu_e})^3 - \left(\sqrt{(M-\sqrt{3T_{he}/\mu_e})^2 + \frac{2\Phi}{\mu_e}}\right)^3\right\}. \quad (4)$$

In this equation, all symbols have the same meaning as in I, viz., $n_{i0}, n_{ce0}, n_{he0}$, are, respectively, the number densities of the ions, and cool and hot electrons, $T_{ce}(T_{he})$ is the temperature of the cool (hot) electrons, $M$ is the Mach number, $\Phi$ is the electrostatic wave potential, and $\mu_e$ is the electron-to-ion mass ratio. The expressions for the second and third derivatives of the unapproximated form of the Sagdeev potential (4) evaluated at $\Phi = 0$, respectively, are given by

$$\left(\frac{d^2V(\Phi)}{d\Phi^2}\right)_{\Phi=0} = \frac{1}{[M^2-3]} + \frac{n_{ce}^0}{\mu_e[M^2-(3T_{ce}/\mu_e)]} + \frac{n_{he}^0}{\mu_e[M^2-(3T_{he}/\mu_e)]} \quad (5)$$

and

$$\left(\frac{d^3V(\Phi)}{d\Phi^3}\right)_{\Phi=0} = \frac{3[M^2+1]}{[M^2-3]^3} - \frac{3n_{ce}^0[M^2+(T_{ce}/\mu_e)]}{\mu_e^2[M^2-(3T_{ce}/\mu_e)]^3} - \frac{3n_{he}^0[M^2+(T_{he}/\mu_e)]}{\mu_e^2[M^2-(3T_{he}/\mu_e)]^3}. \quad (6)$$

As discussed in I, the critical value of the Mach number $M = M_{\text{crit}}$ above which electron-acoustic solitons can exist is given by the higher positive root of Eq. (5), whereas the lower positive root of (5) corresponds to the lower Mach number limit for ion-acoustic solitons. The limitations on the amplitudes of negative potential electron-acoustic solitons are imposed by constraints relating to the number density of the cool electrons or the hot electrons that neither one of these should become complex valued. The number density of the cool electrons (2) will cease to be real valued if $\Phi < \Phi_{\text{min/cool}}$, where $\Phi_{\text{min/cool}} = -\mu_e(M - \sqrt{3T_{ce}/\mu_e})^2/2$ is the limiting value of the potential (negative) of electron-acoustic solitons, the existence of which, in turn, imposes the existence of an upper limit on the Mach number, viz., $M_{\text{max}}$.

Similarly, negative potential solitons can also be limited by the existence of an upper limit on the Mach number which arises because the number density of the hot electrons (3) ceases to be real valued if $\Phi < \Phi_{\text{min/hot}}$, where $\Phi_{\text{min/hot}} = -\mu_e(M - \sqrt{3T_{he}/\mu_e})^2/2$ is now the limiting value of the potential of the negative potential electron-acoustic soliton structures. Whether the existence of an upper limit on the Mach number and amplitude of negative potential electron-acoustic soliton structures is imposed by the number density of the cool or hot electrons becoming complex valued, one has to resort to numerical considerations of $V(\Phi)$ to establish this. The limit on the permitted potentials (either negative or positive) of electron-acoustic soliton structures, which is not related to the number density of any charge particle constituent becoming complex valued, can also be imposed by the occurrence of a double layer. A positive (negative) double layer occurs when a positive (negative) root of $V(\Phi)$, viz., $\Phi_{\text{root}}$ coincides with $(dV(\Phi)/d\Phi) = 0$ at $\Phi = \Phi_{\text{root}}$. It is not obvious from the form of the expression for the Sagdeev potential whether a double layer will or will not occur but this can only be established from numerical considerations of $V(\Phi)$ as discussed in Sec. IV.

### III. MODEL WITH BOLTZMANN HOT ELECTRONS

For the model described in Sec. III in I for which inertia of the hot electrons is not taken into consideration,[6] the number density of the hot electrons which are assumed to be Boltzmann distributed is given by the expression

$$n_{he} = n_{he}^0 \exp\left(\frac{\Phi}{T_{he}}\right). \quad (7)$$

The expressions for the number densities of the adiabatic ions and the adiabatic cool electrons remain unchanged and are given, respectively, by (1) and (2) in Sec. II. The expressions (1), (2) and (7) in Poisson's equation yield for the Sagdeev potential, the expression given by

$$V(\Phi, M) = \frac{1}{6\sqrt{3}}\left\{(M+\sqrt{3})^3 - \left(\sqrt{(M+\sqrt{3})^2 - 2\Phi}\right)^3\right\} - \frac{1}{6\sqrt{3}}\left\{(M-\sqrt{3})^3 - \left(\sqrt{(M-\sqrt{3})^2 - 2\Phi}\right)^3\right\}$$

$$+ \frac{n_{ce}^0}{6\sqrt{3T_{ce}/\mu_e}}\mu_e\left\{(M+\sqrt{3T_{ce}/\mu_e})^3 - \left(\sqrt{(M+\sqrt{3T_{ce}/\mu_e})^2 + \frac{2\Phi}{\mu_e}}\right)^3\right\}$$

$$- \frac{n_{ce}^0}{6\sqrt{3T_{ce}/\mu_e}}\mu_e\left\{(M-\sqrt{3T_{ce}/\mu_e})^3 - \left(\sqrt{(M-\sqrt{3T_{ce}/\mu_e})^2 + \frac{2\Phi}{\mu_e}}\right)^3\right\} + n_{he}^0 T_{he}\left[1 - \exp\left(\frac{\Phi}{T_{he}}\right)\right]. \quad (8)$$

The second and third derivatives of the expression for the Sagdeev potential $V(\Phi)$ (8) evaluated at $\Phi = 0$, respectively, are given by

$$\left(\frac{d^2 V(\Phi)}{d\Phi^2}\right)_{\Phi=0} = \frac{1}{[M^2 - 3]} + \frac{n_{ce}^0}{\mu_e[M^2 - (3T_{ce}/\mu_e)]} - \frac{n_{he}^0}{T_{he}} \quad (9)$$

and

$$\left(\frac{d^3 V(\Phi)}{d\Phi^3}\right)_{\Phi=0} = \frac{3[M^2+1]}{[M^2-3]^3} - \frac{3n_{ce}^0[M^2 + (T_{ce}/\mu_e)]}{\mu_e^2[M^2 - (3T_{ce}/\mu_e)]^3} - \frac{n_{he}^0}{T_{he}^2}. \quad (10)$$

The amplitude of negative potential electron-acoustic soliton structures can still be limited by the constraint that the cool electron number density (2) must be real valued; however, no such restriction is imposed by the hot electrons as their number density is always real as seen from Eq. (7). It will become apparent from our numerical results that electron-acoustic solitons having positive potentials are not supported when the hot electrons are Boltzmann distributed. Positive potential solitons of the electron-acoustic type are only found to be supported by the model of Sec. II when inertial effects of the hot electrons are included and it will be seen later that these are limited only by positive potential double layers.

## IV. NUMERICAL RESULTS AND DISCUSSION

We first consider the model of Sec. II for which inertial effects of all three species, viz., ions, and cool and hot electrons are taken into consideration[18] and investigate the existence of large amplitude electron-acoustic solitons. For the same values of the fixed parameters supporting the occurrence of large amplitude ion-acoustic solitons in Figure 1 in I, here Figure 1(a) depicts the Mach number ranges as a function of $n_{ce0}/n_{i0}$ for which large amplitude electron-acoustic solitons are supported. We have demarcated Figure 1(a) into four different regions, viz., Regions I, II, III, and IV according to the reason for the upper Mach number limit which is unique to each region. The critical value of $M$ for large amplitude electron-acoustic solitons, now, corresponds to the larger[18] of the two positive numerical roots of (5). Choosing a value for $n_{ce0}/n_{i0}$, the existence of electron-acoustic solitons starts at the value of $M$ lying just above the critical value which coincides with a point on (—) in each region, but, depending on the particular fixed value of the number density of the cool electrons $(n_{ce0}/n_{i0})$, terminates at the value of $M$ lying just below one of the curves, viz., $(\cdots)$ which defines Region I, (- - -) which defines Region II, $(-\cdot-)$ which defines Region III, or $(-\cdot\cdot\cdot-)$ which defines Region IV.

Region I includes values of the Mach number which lie between the lower $(M_{crit})$ and upper $(M_{max})$ Mach number limiting curves, respectively, denoted by (–) and $(\cdots)$ in Figure 1(a) for $0.05 \le n_{ce0}/n_{i0} \le 0.174$. This corresponds to a region in parameter space where electron-acoustic solitons having negative potentials are supported, but, where the existence of an upper limit on the Mach number, viz., $M_{max}$, is imposed by the constraint that the cool electron number density (2) must remain real valued, i.e., $\Phi \geq \Phi_{\min/\text{cool}} = -\mu_e(M - \sqrt{3T_{ce}/\mu_e})^2/2$ as discussed in Sec. II. With increasing values of $M$, electron-acoustic solitons having negative potentials become stronger as can be seen in Figure 2 for $n_{ce0}/n_{i0} = 0.1$ which shows plots of the Sagdeev potential, $V(\Phi)$, versus $\Phi$. The increase in soliton amplitudes with increasing $M$ will not occur indefinitely because electron-acoustic solitons will cease to exist once $\Phi_{\min/\text{cool}}$ is reached. The upper limiting value of the Mach number is encountered where $V(\Phi_{\min/\text{cool}}) = 0$. This situation corresponds to the curve represented by (– $\cdot$ –) in Figure 2 where the upper limit on the Mach number $M = M_{\max} = 70.81841$ coincides with the minimum permitted value of $\Phi$, viz., $\Phi = \Phi_{\min/\text{cool}} = -1.09454$. The upper limiting plot of the Sagdeev potential for $M = M_{\max}$ in Figure 2 does not yield a valid soliton solution. It is also clear from Figure 2 that a soliton does not occur for a higher value of $M$, viz., $M = 72$ (– $\cdot\cdot\cdot$ –) which exceeds $M_{\max}$, in which case, the cool electron number density (2) becomes complex valued. A similar argument applies to solitons which occur in Region II and these will also cease to exist for $M \geq M_{\max}$ because the number density of the hot electrons (3) becomes complex valued.

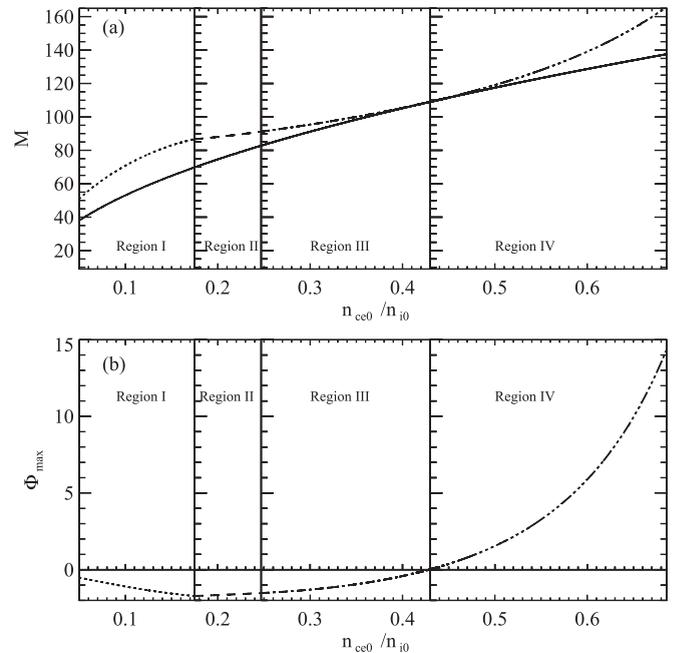

FIG. 1. (a) Existence domains of negative and positive potential electron-acoustic solitons shown as a function of the normalized cool electron number density. The curve (–) denotes $M_{crit}(n_{ce0}/n_{i0})$, $(\cdots)$ denotes maximal $M$ values beyond which the cool electron number density (2) is not real valued, (- -) denotes maximal $M$ values beyond which the hot electron number density (3) is not real valued, (- $\cdot$ -) denotes Mach numbers for which negative double layers occur and (- $\cdot\cdot\cdot$ -) denotes Mach numbers supporting positive double layers. (b) Limiting values of the potentials corresponding to the upper $M$ limits for solitons in (a). Region I (Region II) shows potential limits beyond which the cool electron number density (2) (the hot electron number density (3)) becomes complex valued. In Region III (Region IV), negative (positive) double layer potentials are shown. The fixed parameters are $\mu_e = 1/1836$, $T_{ce}/T_i = 0.01$, and $T_{he}/T_i = 5$.

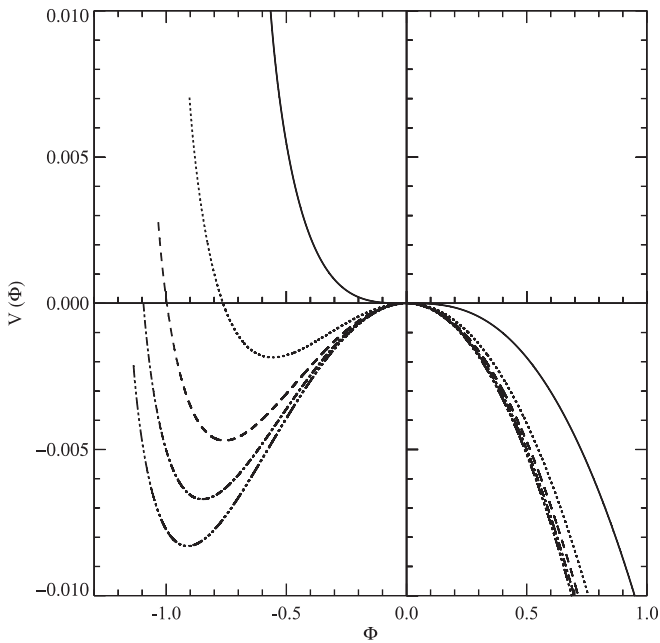

FIG. 2. Sagdeev potential profiles for $M = 53.07343$ (–), 65 ($\cdots$), 69 (– –), 70.81841 (– $\cdot$ –), and 72 (– $\cdot\cdot\cdot$ –). The fixed parameters are $\mu_e = 1/1836$, $T_{ce}/T_i = 0.01$, $T_{he}/T_i = 5$, and $n_{ce0}/n_{i0} = 0.1$.

Following the ideas in Ref. 23, the upper Mach number limiting curve shown as ($\cdots$) in Figure 1(a) has been generated by numerically solving $V(\Phi_{\min/\text{cool}}) = 0$ for $M$. The variation of $\Phi_{\min/\text{cool}}$ with $n_{ce0}/n_{i0}$ for negative potential electron-acoustic solitons which occur in Region I is denoted by the curve ($\cdots$) in Figure 1(b). The values $\Phi_{\min/\text{cool}}$ in Figure 1(b) have been evaluated using the upper $M$ limits shown in Figure 1(a) for solitons which occur in Region I.

We again refer to Figure 2 from which it can be inferred that lower and upper limiting values of the Mach number exist for electron-acoustic solitons as is clearly realized from plots of the Sagdeev potential (4) for different values of $M$ for $n_{ce0}/n_{i0} = 0.1$. The fixed value $n_{ce0}/n_{i0} = 0.1$ lies in Region I which is bounded from above by ($\cdots$) in Figure 1(a). The lower limiting curve (–) in Figure 2 corresponds to the critical value of the Mach number $M = 53.07343$ for $n_{ce0}/n_{i0} = 0.1$. The upper limiting curve denoted by (– $\cdot$ –) in Figure 2 for the upper limiting value $M = 70.81841$ coincides with the value $\Phi_{\min/\text{cool}}$, which is the minimum permitted value of the potential (negative) for which the cool electron number density (2) is still real valued. The value $\Phi_{\min/\text{cool}}$ corresponding to $n_{ce0}/n_{i0} = 0.1$ is easily realised from plots of the Sagdeev potential for $M$ values up to the upper limiting value $M = 70.81841$ in Figure 2 and $\Phi_{\min/\text{cool}}$ is precisely the limiting value of the negative roots of $V(\Phi)$. We find this value to be $-1.09454$ which coincides with the negative root of the upper limiting plot of $V(\Phi)$ denoted by the curve (– $\cdot$ –) in Figure 2. The value $-1.09454$ for $\Phi_{\min/\text{cool}}$ coincides with the point corresponding to $n_{ce0}/n_{i0} = 0.1$ on the curve denoted by ($\cdots$) (Region I) in Figure 1(b).

For plasmas with higher concentrations of cool electrons, now, lying in the range $0.175 \leq n_{ce0}/n_{i0} \leq 0.246$ corresponding to Region II in Figure 1(a), the upper limit on the Mach number restricting the occurrence of negative potential electron-acoustic solitons now coincides with the limiting value of the negative potential which is imposed by the constraint that the number density of the hot electrons (3) must remain real valued, i.e., $\Phi \geq \Phi_{\min/\text{hot}} = -\mu_e(M - \sqrt{3T_{he}/\mu_e})^2/2$. Limiting values of the negative potential, viz., $\Phi_{\min/\text{hot}}$ such that the number density of the hot electrons (3) is no longer real valued for $\Phi < \Phi_{\min/\text{hot}}$, coincide with points which lie on the curve denoted by (- - -) in Figure 1(b). The upper $M$ limits for Region II which lie on the curve (- - -) in Figure 1(a) were obtained by numerically solving $V(\Phi_{\min/\text{hot}}) = 0$ as a function of $n_{ce0}/n_{i0}$. The values $\Phi_{\min/\text{hot}}$ shown for Region II in Figure 1(b) were calculated using the corresponding upper Mach number limits shown for this region in Figure 1(a).

For the fixed value $n_{ce0}/n_{i0} = 0.2$ which lies in Region II which is bounded from above by (- - -) in Figure 1(a), the negative root of the upper limiting plot of the Sagdeev potential denoted by (– $\cdot$ –) in Figure 3 corresponding to the value $M = 87.99353$, coincides with the limiting value of the potential $\Phi_{\min/\text{hot}}$. The value $-1.65509$ for $\Phi_{\min/\text{hot}}$ is clearly apparent from Figure 3 where $\Phi_{\min/\text{hot}}$ is the limiting value of the negative roots of the Sagdeev potentials depicted in the figure. This value $-1.65509$ which is the limit on $\Phi$ coincides with the point corresponding to $n_{ce0}/n_{i0} = 0.2$ on (- - -) in Region II in Figure 1(b).

Considering now higher concentrations of the cool electrons such that $0.247 \leq n_{ce0}/n_{i0} < 0.43$, there is a switch to a region in parameter space where negative double layers are found to limit the occurrence of electron-acoustic solitons having negative potentials. This is shown as Region III in Figure 1(a) which includes values of the Mach number which lie just above the curve (–) but terminates just below the curve denoted by (– $\cdot$ –). For the fixed value $n_{ce0}/n_{i0} = 0.3$ which coincides with soliton Region III which is bounded from above by (– $\cdot$ –) in Figure 1(a), the existence of negative potential electron-acoustic solitons terminates at the value of $M$ which lies on (– · –) for which the occurrence of a negative potential double layer is supported. This is clearly seen in Figure 4, where for $n_{ce0}/n_{i0} = 0.3$, the upper limiting plot of the Sagdeev potential, now, coincides with the occurrence of a negative potential double layer as demonstrated by the behaviour of the curve denoted by (– · –) for $M = 95.417998$. It is seen in Figure 4 that when the double layer Mach number limit is exceeded, electron-acoustic solitons are no longer possible as shown for $M = 95.7$ (– · · –).

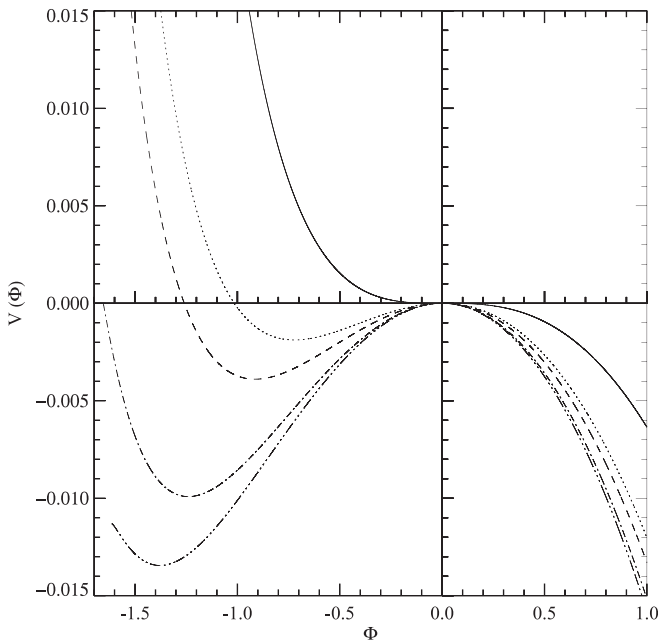

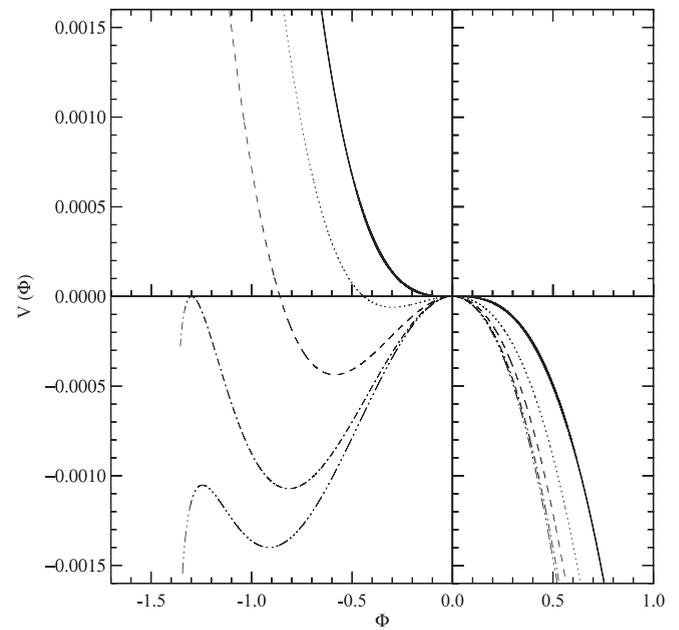

FIG. 3. Sagdeev potential profiles for $M = 74.59170$ (—), 83 ($\cdots$), 85 (– –), 87.99353 (– · –), and 89 (– · · –). The fixed parameters are $\mu_e = 1/1836$, $T_{ce}/T_i = 0.01$, $T_{he}/T_i = 5$, and $n_{ce0}/n_{i0} = 0.2$.

FIG. 4. Sagdeev potential profiles for $M = 91.16449$ (—), 93 ($\cdots$), 94.5 (– –), 95.417998 (– · –), and 95.7 (– · · –). The fixed parameters are $\mu_e = 1/1836$, $T_{ce}/T_i = 0.01$, $T_{he}/T_i = 5$, and $n_{ce0}/n_{i0} = 0.3$.

Reverting to Figure 1(a), but focusing now on Region IV which corresponds to $n_{ce0}/n_{i0} > 0.43$, upper limiting values of $M$ which lie on the curve (– · · –) in Figure 1(a) now coincide with the occurrence of *positive* potential double layers, which limits the region in parameter space where electron-acoustic solitons having *positive* potentials are supported. The switch in polarity of electron-acoustic solitons from *negative*, for $n_{ce0}/n_{i0} < 0.43$, to *positive*, for $n_{ce0}/n_{i0} > 0.43$, is clear from Figure 1(b) which depicts the maximum potentials of negative (positive) potential double layers corresponding to the negative (positive) potential soliton regions (Regions III and IV) depicted in Figure 1(a). The switch in polarity of electron-acoustic solitons which we observe is consistent with the change of sign of $C_3(M_{\text{crit}})$ (Eq. (17) in I) from negative for $n_{ce0}/n_{i0} < 0.43$ to positive for $n_{ce0}/n_{i0} > 0.43$ (the small amplitude soliton solution was given as Eq. (15) in I).

It is clear from Figure 1(b) that $n_{ce0}/n_{i0} = 0.43$ corresponds to a neutral point ($\Phi = 0$), where the separation between Region III (negative potential solitons limited by negative double layers) and Region IV (positive potential solitons limited by positive double layers) occurs. The particular value of $n_{ce0}/n_{i0} = 0.43$ at which the switch in polarity of the nonlinear electron-acoustic structures (solitons and double layers) is observed to occur is in agreement with the findings for electron-acoustic solitons in Lakhina *et al.*,[18] wherein the switch in polarity of electron-acoustic solitons was also reported to occur at the value $n_{ce0}/n_{i0} = 0.43$, but this was calculated for a lower fixed value of the temperature of the hot electrons, viz., $T_{he}/T_i = 1$ rather than the value $T_{he}/T_i = 5$ used by us to generate Figure 1. Furthermore, $n_{ce0}/n_{i0} = 0.43$ also corresponds to the crossover point from negative to positive potential electron-acoustic soliton regions limited by double layers for the $\gamma = 3$ case in the study by Cattaert *et al.*[14] wherein the fluid-dynamic paradigm approach was used. Here we have established that positive potential double layers limit the existence of positive potential electron-acoustic solitons, considerably broadening the scope of the findings in Ref. 18 since there is no mention of double layers in Ref. 18. We have terminated Figure 1 at $n_{ce0}/n_{i0} = 0.686$, since, our findings reveal that positive double layers cease to exist for cool electron number densities which exceed $n_{ce0}/n_{i0} = 0.686$. For $n_{ce0}/n_{i0} = 0.6$, the positive double layer corresponding to a plot of $V(\Phi)$ denoted by the curve (– · –) for $M = 139.04997$ in Figure 5, is seen to limit the occurrence of electron-acoustic solitons having *positive* potentials. The existence regions depicted in Figure 1(a) for large amplitude electron-acoustic solitons, where, the amplitude restrictions are seen to switch from number density constraints pertaining to the negatively charged cool species and then the negatively charged hot species, followed by the occurrence of negative double layers and finally positive double layers, is very similar to those found for large amplitude electron-acoustic solitons by Cattaert *et al.*[14] and dust-acoustic solitons by Verheest *et al.*[23] To conclude our discussion of Figure 1, we are pleased to remark that the parameter ranges over which large amplitude electron-acoustic solitons are seen to occur in Figure 1, but, also later in Figure 6, are all well within the permitted ranges, viz., $f = n_{ce0}/n_{i0} < 0.8$

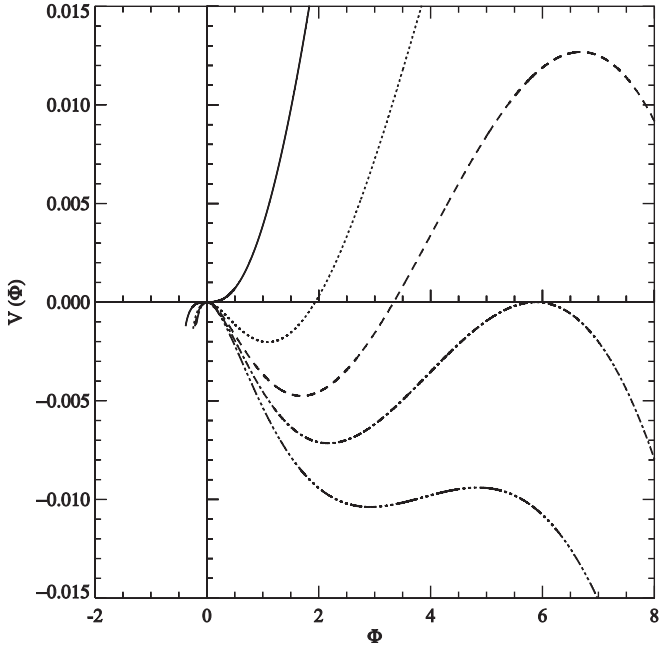

FIG. 5. Sagdeev potential profiles for $M = 128.65457$ (–), $136$ ($\cdots$), $138$ (– –), $139.04997$ (– $\cdot$ –), and $140$ (– $\cdot\cdot$ –). The fixed parameters are $\mu_e = 1/1836$, $T_{ce}/T_i = 0.01$, $T_{he}/T_i = 5$, and $n_{ce0}/n_{i0} = 0.6$.

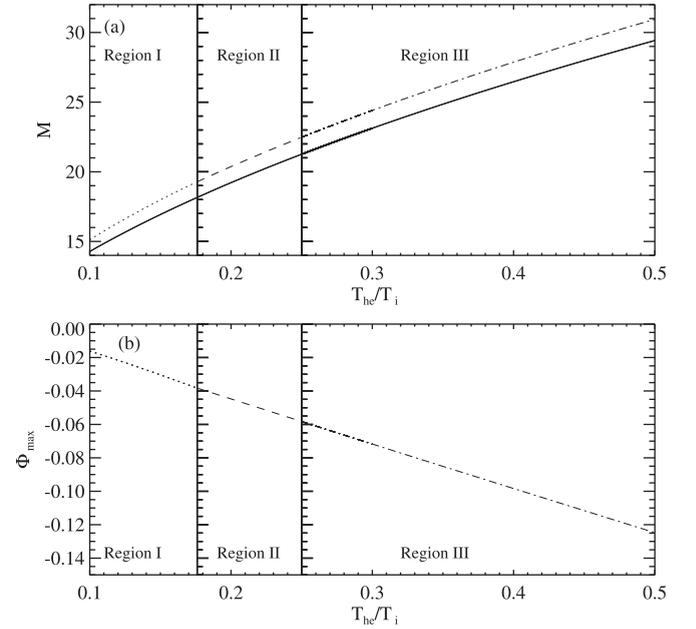

FIG. 6. (a) Existence domains of negative potential electron-acoustic solitons shown as a function of the normalized hot electron temperature $T_{he}/T_i$. The curve (–) denotes $M_{\text{crit}}(T_{he}/T_i)$, ($\cdots$) denotes maximal $M$ values beyond which the cool electron number density (2) is not real valued, (- -) denotes maximal $M$ values beyond which the hot electron number density (3) is not real valued, and (– $\cdot$ –) denotes Mach numbers for which negative double layers occur. (b) Limiting values of the negative potentials which correspond to the upper $M$ limits for solitons in (a). Regions I and II, respectively, show the potential limits beyond which the number density of the cool electrons (2) and number density of the hot electrons (3) are no longer real valued, whereas the potentials shown for Region III are negative double layer potentials. The fixed parameters are $\mu_e = 1/1836$, $T_{ce}/T_i = 0.01$, and $n_{ce0}/n_{i0} = 0.3$.

and $T_h/T_c > 1$ for only weakly damped linear electron-acoustic waves.[25,26]

Fixing the number density of the cool electrons, viz., $n_{ce0}/n_{i0} = 0.3$, we next investigate existence domains of large amplitude electron-acoustic solitons as a function of the normalized temperature of the hot electrons, viz., $T_{he}/T_i$. In Region I ($0.1 \leq T_{he}/T_i \leq 0.175$) of Figure 6(a) negative potential electron-acoustic solitons are supported, and upper limits on the Mach number, $M_{\max}$, and the soliton potential $\Phi$ are imposed by the constraint that the number density of the cool electrons (2) must remain real valued. The upper $M$ limits for solitons which occur in Region I are shown as ($\cdots$) in Figure 6(a). The variation of $\Phi_{\min/\text{cool}}$ with $T_{he}/T_i$ is denoted by the curve ($\cdots$) (Region I) in Figure 6(b). The intermediate region in parameter space which spans $0.176 \leq T_{he}/T_i \leq 0.249$ (Region II) in Figure 6(a) also supports the existence of negative potential electron-acoustic solitons, but, here, the limiting value on the potential (negative) and the existence of an upper limit on the Mach number (- - -) are imposed by the constraint that the number density of the hot electrons (3) has to remain real valued. For electron-acoustic solitons which occur in Region II, limiting values of the negative potential, viz., $\Phi_{\min/\text{hot}}$, lie on the curve denoted by (- - -) in Figure 6(b). In Region III ($0.25 \leq T_{he}/T_i \leq 0.5$) depicted in Figure 6(a), large amplitude electron-acoustic solitons having negative potentials are limited by negative potential double layers. The upper $M$ limits in Region III which lie on the curve denoted by (– $\cdot$ –) in Figure 6(a) give rise to negative potential double layers. The admissible soliton Mach number ranges depicted in Figure 6(a) are seen to widen but only very slightly with increasing values of the ratio $T_{he}/T_i$. Furthermore, the magnitudes of the negative double layer potentials (– $\cdot$ –) shown for Region III in Figure 6(b) reveal that double layers become stronger with increasing disparity between the temperatures of the cool and hot electron components (increasing $T_{he}/T_i$ for a fixed value of $T_{ce}/T_i$).

Finally, we focus on comparing our results for the model of Sec. II which considers inertial effects of the hot electrons[18] with the model of Mace *et al.*[6] which neglects inertial effects of the hot electrons as discussed in Sec. III. So as not to overload this paper, here we do not include any of our results for the model of Sec. III, but, merely provide a brief summary of our findings. When inertial effects of the hot electrons are neglected, our results are consistent with the findings in Ref. 6 in that only negative potential electron-acoustic solitons are possible. Furthermore, we found that the limiting values on the potential (negative) and Mach number, $M_{\max}$, were imposed by the constraint that the number density of the cool electrons must remain real valued, i.e., $\Phi \geq \Phi_{\min/\text{cool}}$. No limitation on the potential or the Mach number was imposed by the hot electrons as their number density remains real valued for any value of the potential. As expected, we did not find any negative potential double layers or positive potential electron-acoustic solitons (limited by positive potential double layers) as a minimum of two inertial electron constituents are necessary.[14,18,21] Furthermore, for the model for which the inertia of the hot electrons is neglected, we have identified a region in parameter space corresponding to $n_{ce0}/n_{i0} > 0.56$ where large amplitude electron-acoustic solitons having negative potentials are found not to have an upper Mach number limit.

It is quite interesting how the change in polarity of electron-acoustic solitons from negative to positive is induced and how these solitons which occur in adjacent regions of parameter space are limited by double layers (both polarities are supported) by including hot electron inertia[18] as opposed to neglecting the inertia of the hot electrons[6] in the model.

## V. CONCLUSIONS

We have investigated the existence of large amplitude electron-acoustic solitons for a three-component plasma composed of ions, and cool and hot electrons. We considered two models, namely, where all species are treated as mobile[18] and where inertia and pressure of the ions and cool electrons are taken into account but the inertia of the hot electrons is neglected.[6] The regions in parameter space supporting the existence of large amplitude electron-acoustic solitons have been identified and are presented here for the three-component model which includes inertial effects for all species which are assumed to be adiabatic consistent with the model of Lakhina *et al.*[18]

Our primary focus was to first identify why upper Mach number limits exist for large amplitude electron-acoustic solitons, and then explicitly calculate these upper Mach number limits for much broader regions in parameter space than those considered in Ref. 18. In doing so, we found not only parametric regions where negative potential double layers limit the occurrence of the negative potential solitons but also other regions where positive potential double layers were found as upper limits on the Mach number ranges supporting electron-acoustic solitons having positive potentials.

For the model which includes inertia of the hot electrons,[18] starting from the smallest concentrations of the cool electrons ($0.05 \leq n_{ce0}/n_{i0} \leq 0.174$), we initially obtain negative potential electron-acoustic solitons where the constraint on the potential and the existence of an upper limit on the Mach number, viz., $M_{max}$, arise from the constraint that the number density of the cool electrons must remain real valued. For higher concentrations of the cool electrons, there is a switch to a region in parameter space where the number density of the hot electrons must remain real valued imposes the existence of the upper $M$ limit, $M_{max}$, restricting the amplitudes of the negative potential electron-acoustic solitons occurring for $0.174 < n_{ce0}/n_{i0} \leq 0.246$. Increasing the number density of the cool electrons further such that $0.246 < n_{ce0}/n_{i0} < 0.43$, we obtain negative potential electron-acoustic solitons which are limited by double layers (negative potential). Finally, the highest concentrations of the cool electrons spanning $0.43 < n_{ce0}/n_{i0} \leq 0.686$ also support the occurrence of electron-acoustic solitons, but these are now found to have positive potentials and are limited only by positive potential double layers. The particular value $n_{ce0}/n_{i0} = 0.43$ at which we have observed the switch in polarity of electron-acoustic solitons to occur from negative ($n_{ce0}/n_{i0} < 0.43$) to positive ($n_{ce0}/n_{i0} > 0.43$) for our chosen fixed value $T_{he}/T_i = 5$ is consistent not only with the findings in Lakhina *et al.*[18] for the lower value $T_{he}/T_i = 1$ but also Cattaert *et al.*[14] for the $\gamma = 3$ case. We found that positive potential double layers cease to exist beyond $n_{ce0}/n_{i0} = 0.686$.

Reverting to the model of Mace *et al.*[6] for which the inertia and pressure of the ions and cool electrons are retained but the inertialess hot electrons are assumed to be Boltzmann distributed, our results reveal that only negative potential electron-acoustic solitons can be supported, which is consistent with the findings for electron-acoustic solitons having small and large amplitudes in Ref. 6. For a wide range of values of $n_{ce0}/n_{i0}$, the upper Mach number limiting the occurrence of negative potential electron-acoustic structures coincides only with the limiting value of the potential (negative) for which the cool electron number density is still real valued, never with the occurrence of a double layer (negative). We have established that for cool electron concentrations which exceed $n_{ce0}/n_{i0} = 0.56$, the Mach number ranges supporting the existence of negative potential electron-acoustic solitons appear not to have upper bounds when the hot electrons are Boltzmann distributed.

## ACKNOWLEDGMENTS

S.V.S. and R.B. would like to thank the Department of Science and Technology, New Delhi, India and NRF of South Africa, respectively, for the financial support. The work was done under Indo-South Africa Bilateral Project "Linear and nonlinear studies of fluctuation phenomena in space and astrophysical plasmas." G.S.L. thanks the Indian National Science Academy, New Delhi, India, for the support under the Senior Scientist scheme.